\documentclass[journal,twocolumn,10pt,twoside]{IEEEtran}
\usepackage[dvips,final]{graphicx}
\usepackage{latexsym}
\usepackage{amssymb}
\usepackage{amsmath}
\usepackage{amsthm}
\usepackage{bm}
\usepackage{array}
\usepackage{balance}
\usepackage{cite}
\usepackage{subfig}

\newtheorem*{example}{Example}

\title{Decoding Delay Minimization in Inter-Session Network Coding}

\author{Eirina~Bourtsoulatze,~\IEEEmembership{Student Member,~IEEE,}
	     Nikolaos~Thomos,~\IEEEmembership{Member,~IEEE,} 
              and~Pascal~Frossard,~\IEEEmembership{Senior Member,~IEEE} %
\thanks{E. Bourtsoulatze and P. Frossard are with the Signal Processing Laboratory 4 (LTS4), Ecole Polytechnique F\'{e}d\'{e}rale  de Lausanne (EPFL), Lausanne, Switzerland (e-mail: eirina.bourtsoulatze@epfl.ch; pascal.frossard@epfl.ch). N. Thomos is with the Signal Processing Laboratory 4 (LTS4), Ecole Polytechnique F\'{e}d\'{e}rale  de Lausanne (EPFL), Lausanne, Switzerland, and the Communication and Distributed Systems laboratory (CDS), University of Bern, Bern, Switzerland (e-mail: nikolaos.thomos@epfl.ch).}%
\thanks{This work has been supported by the Swiss National Science Foundation under grants 200021-118230, 200021-138083, PZ00P2-121906 and PZ00P2-137275.}}

\begin{document}

\maketitle


\begin{abstract}
Intra-session network coding has been shown to offer significant gains in terms of achievable throughput and delay in settings where one source multicasts data to several clients. In this paper, we consider a more general scenario where multiple sources transmit data to sets of clients and study the benefits of inter-session network coding, when network nodes have the opportunity to combine packets from different sources. In particular, we propose a novel framework for optimal rate allocation in inter-session network coding systems. We formulate the problem as the minimization of the average decoding delay in the client population and solve it with a gradient-based stochastic algorithm. Our optimized inter-session network coding solution is evaluated in different network topologies and compared with basic intra-session network coding solutions. Our results show the benefits of proper coding decisions and effective rate allocation for lowering the decoding delay when the network is used by concurrent multicast sessions. 
\end{abstract}



\section{Introduction}
\label{sec:Intro}

The recent advances in adhoc and overlay networks have largely contributed to the development of network coding algorithms \cite{Ahlswede00,LinearNC03}. These networks are characterized by a variety of resources, and data is delivered from multiple sources to sets of clients through several overlapping paths. This creates network coding opportunities with possibly large gains in terms of throughput, delay or error robustness. However, the assumption posed by many network coding systems is that the network is utilized by a single source \cite{ChouPractical03,WangR207}. This is quite restricting as multiple sessions coexist over the shared network resources in most realistic scenarios. Time sharing or optimal bandwidth allocation of the overlapping paths can be considered in order to make effective use of shared resources. Alternatively, data from different sources can be combined in network nodes with so-called inter-session network coding algorithms \cite{Dougherty05,LinearInterNC}. These algorithms have gained quite a lot of interest recently as they theoretically permit to obtain throughput and delay gains compared to multiplexing solutions. This is the type of systems that we consider in this paper. 

\begin{figure}[t]
	\begin{center}
		~\includegraphics[width=0.45\textwidth]{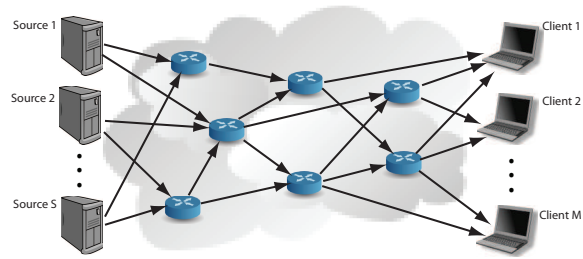}~
	\end{center}
	\caption{Illustration of multi-source data delivery on an overlay network. }
	\label{fig:network}
\end{figure}

More precisely, we consider lossy networks with significant path diversity where multiple sources attempt to simultaneously deliver data to different clients, as depicted in Fig.\ref{fig:network}. Inter-session network coding is implemented in network nodes in order to exploit the path diversity and reduce the decoding delay without the need for explicit scheduling. Specifically, we employ inter-session network coding based on randomized linear network coding \cite{RandomizedNC03}. In more details, the sources encode the data with randomized linear coding and send the coded data to the downstream nodes. The packets arriving at a node are first stored and when a transmission opportunity occurs a network coded packet is sent to a child node.

The application of inter-session network coding is not, however, a trivial task as random mixing of packets from different data flows may result in unacceptably large decoding delays and waste of network resources. We therefore formulate an optimization problem to select the appropriate coding operations in network nodes, such that the average decoding delay is minimized among clients. Our target is to determine the optimal coding strategy in the intermediate network nodes and the rate allocation among the intra- and inter-session network coded flows. Experimental evaluation shows the validity of our model. The application of inter-session network coding leads to lower decoding delay in several network topologies, or to the same performance as  intra-session network coding in the least favorable configurations. To the best of our knowledge, our scheme is the only scheme in the literature that provides a complete analytical inter-session network coding framework that is easily extensible to any arbitrary number of concurrent sources. It further shows that an effective combination of intra- and inter-session network coding permits to reach small decoding delays even in challenging network settings.


The design of efficient inter-session network coding schemes \cite{KhreishahJSAC09,ErezJSAC09} is largely an open problem in the literature. Combinations of different flows may decrease the network goodput and increase the decoding delays since a larger number of packets has to be collected before the decoding of one of the flows becomes feasible. Most of the schemes in the literature apply inter-session network coding \cite{KhreishahJSAC09,WangAllerton07,SefPV09} for wireless scenarios and exploit transmission interferences and overhearing of packets intended to other clients \cite{COPE}. Inter-session network coding in binary field is examined in \cite{KhreishahJSAC09}. This scheme incorporates COPE \cite{COPE}, an opportunistic network coding scheme for wireless multi-hop networks, and applies coding only between two wireless sessions. A distributed algorithm is provided for joint coding, scheduling and rate control. Since the perfect scheduling is not an easy task, the performance loss due to imperfect scheduling is also examined.  COPE is also employed by \cite{KimJSAC09} for star shaped networks where a compromise between the transmission rate and the overhear rate is found for increasing network coding chances. This adaptive scheduling scheme has high complexity and achieves only marginal throughput gains. The work in \cite{WangAllerton07} proposes to control the overlap of transmission paths to take benefit of inter-session network coding. The authors in \cite{SefPV09} decouple the streaming over wireless networks into two independent problems, namely rate-control and scheduling. This introduces some delays in order to increase the coding opportunities for wireless streaming scenarios. A distributed algorithm for the multi-commodity transmission in network with two sources is developed in \cite{ErezJSAC09}.  Inter-session network coding is used along with backpressure routing and the resulting scheme has complexity comparable to that of parallel multi-commodity flow problem (without network coding).

The extension of inter-session network coding to multi-hop wired scenarios is challenging and until now only few works have addressed it \cite{LinearInterNC,Wu06,KhreishanINFOCOM08}. Specifically, a heuristic algorithm has been proposed in \cite{LinearInterNC} for data multicasting. The sessions are divided into groups according to a parameter that measures the overlap among the different sessions and inter-session network coding is implemented only inside these groups. The results show increased throughput and reduced bandwidth consumption; however, the delays and packet losses are not addressed. The work in \cite{Wu06} follows a pollution-free approach, where inter-session network coding is restricted only to the sources that the clients want to receive. Each network link is split into conceptual links that carry all possible combinations of flows and the clients connect to the links containing the sources they are interested in. When each client is interested in a single source, then the system performs intra-session network coding only. Decentralized pairwise network coding is proposed in \cite{KhreishanINFOCOM08} for optimized distributed rate control. Fairness is taken into account for achieving socially optimal behavior and noticing throughput gains. Although this approach is interesting, it is not obvious to understand if the results can be generalized to larger topologies and how the system's performance scales with the number of sources.

The closest work to our study is the scheme considered in \cite{SefI2NC}, where the parallel application of intra- and inter-session network coding for introducing redundancy at intermediate nodes is considered. The benefits of joint consideration of intra- and inter-session network coding become clear in this work, and confirm our findings.  However, the work in \cite{SefI2NC} studies a wireless scenario and different challenges become apparent like opportunistic listening, transmission interference {\em etc}. Gains in terms of throughput and resilience to network losses are presented in simple cross topologies and for bitwise XOR coding ({\em i.e.}, COPE-based solutions). Our work addresses the delay minimization by effective coding solutions in generic wired multi-hop networks.

The rest of the paper is organized as follows. In Section~\ref{sec:interNC}, we provide a brief description of the intra- and inter-session network coding techniques, and give an example that illustrates the benefits of applying inter-session network coding in multi-source network scenarios. We then describe in Section~\ref{sec:system} the architecture of our system, which employs inter-session network coding for data delivery. In Section \ref{sec:ratealloc} we formulate our novel optimization problem for optimal rate allocation in order to achieve minimal decoding delay. Finally, in Section~\ref{sec:evaluation}, we evaluate the performance of the proposed scheme in different settings and Section~\ref{sec:conclusions} concludes the paper.


\section{Inter-session network coding}
\label{sec:interNC}

Network coding has been proposed as an alternative to the traditional routing and scheduling algorithms deployed for network data delivery. The main idea of network coding is to allow intermediate network nodes to process packets and transmit combinations of incoming packets\footnote{In this paper, we use the term ``packet" or ``symbol" interchangeably to denote the elementary entity in coding operations.}. These simple operations performed in the network nodes result in an increased throughput, which potentially implies a decrease in the delay required for data delivery. Network coding also enhances robustness to packet losses.

\begin{figure*}[t]
	\begin{center}
			\subfloat[]{\label{fig:examplerouting}\includegraphics[width=0.4\textwidth]{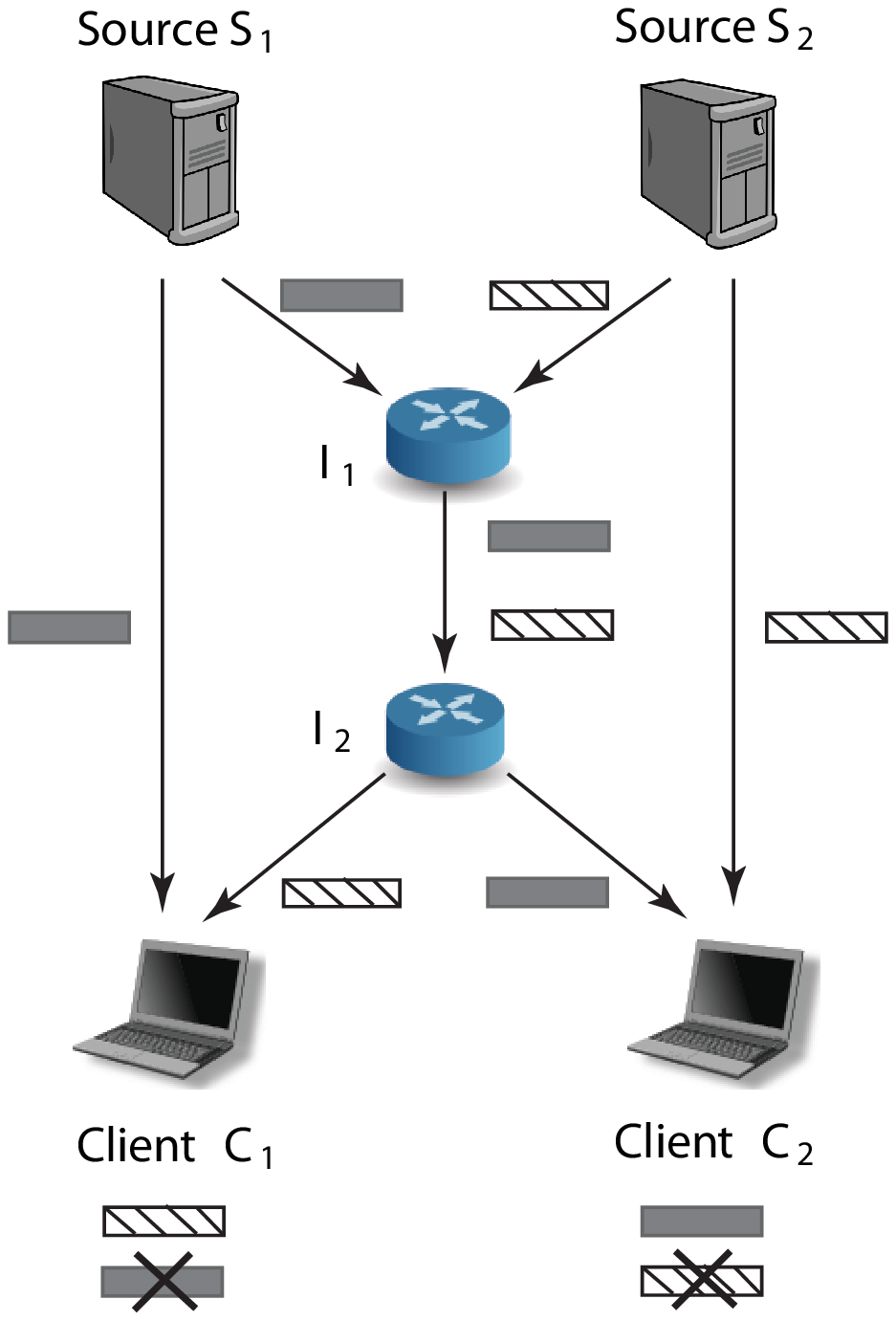}}
			\subfloat[]{\label{fig:exampleInterNC}\includegraphics[width=0.4\textwidth]{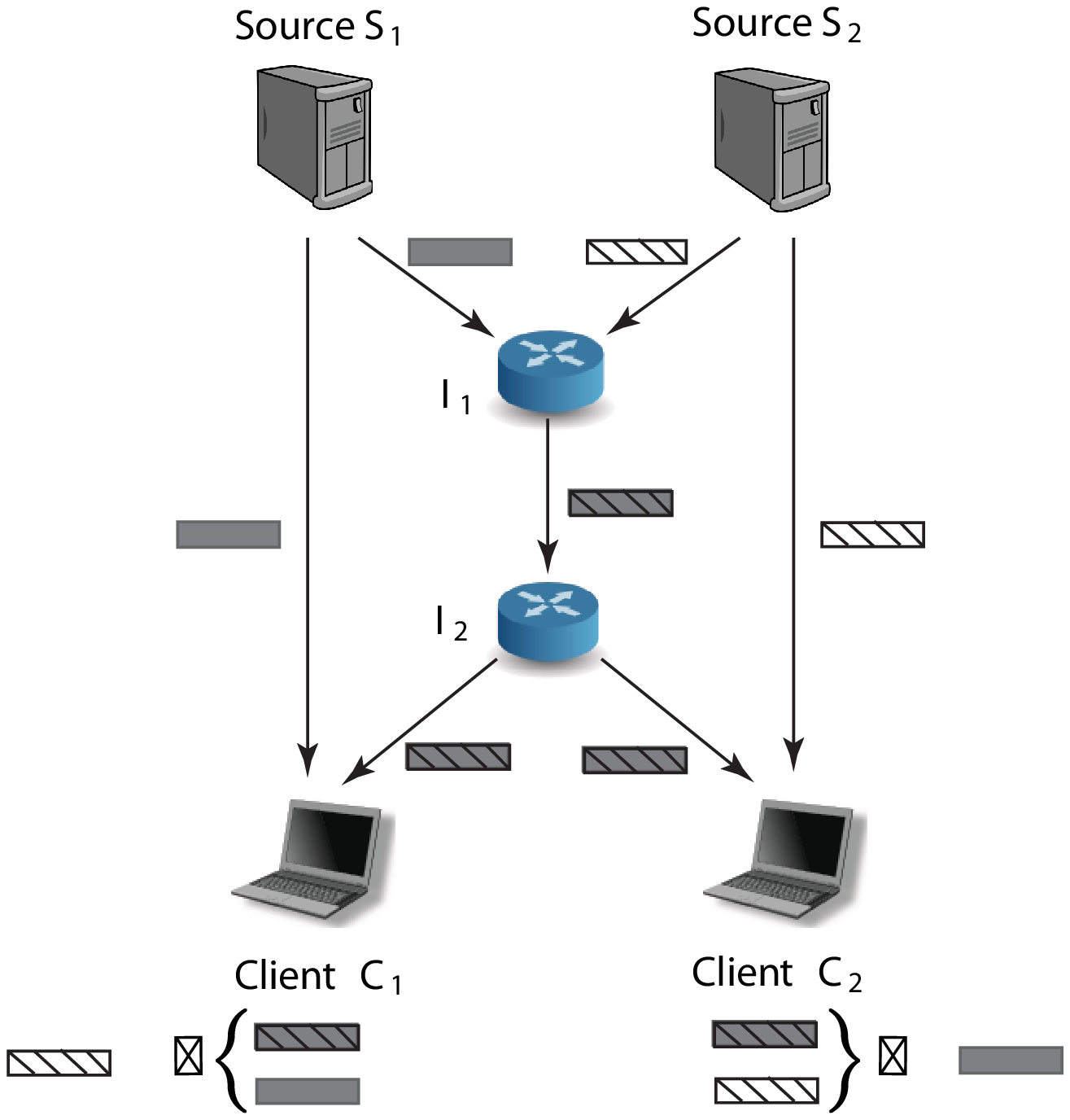}}
	\end{center}
	\caption{Illustration of (a) routing and (b) inter-session network coding in a butterfly network with two unicast sessions.\label{fig:butterfly}}
\end{figure*}

The gain obtained by applying an inter-session network coding scheme can be illustrated by the classical example depicted in Fig.~\ref{fig:butterfly}. In this example, we consider a butterfly network with two simultaneous unicast sessions. The network is composed of two sources $S_1$ and $S_2$, and two clients $C_1$ and $C_2$. The client $C_1$ is interested in receiving the packets from the source $S_2$ and the client $C_2$ is interested in receiving the packets from source $S_1$. The intermediate nodes $I_1$ and $I_2$ can either act as relay nodes by simply forwarding the incoming packets according to a certain scheduling policy or perform inter-session network coding on the input messages. All links have a capacity of one packet per time slot. In this topology, each session has only one path from the source to the client and paths are overlapping in the segment between the nodes $I_1$ and $I_2$. Hence, when the node $I_1$ simply stores and forwards the incoming packets (Fig.~\ref{fig:examplerouting}), the link $(I_1,I_2)$ becomes a bottleneck since it can only support the transmission of one packet per time slot. Depending on the scheduling policy that node $I_1$ adopts, ({\em i.e.}, which packet will be forwarded first), one of the clients experiences an extra delay with respect to the other client. Moreover, in this case, the additional bandwidth provided by the links $(S_1,C_1)$ and $(S_2,C_2)$ remains unexploited, since the information delivered on these links is useless for the clients. It is obvious that, in this simple topology, routing in the intermediate nodes leads to a suboptimal solution and a rather conservative utilization of the network resources. 

On the contrary, if the intermediate node $I_1$ implements inter-session network coding and combines the incoming packets from the two sessions as shown in Fig.~\ref{fig:exampleInterNC}, the performance of the network can be improved. In particular, if the node $I_1$ implements inter-session network coding, the network coded packet, that contains information of both sources reaches both clients at the same time. Both clients are able to extract the information that they are interested in by solving a simple system of equations which is formed of the network coded packet and the packet that the clients have received directly from the sources. Even though the clients are not interested in the information contained in the uncoded packet, this packet is still useful for decoding the requested data. The network throughput is improved with inter-session network coding and both clients experience the same delay. In addition, all network resources are fully exploited in this case.

We provide now more details about the linear network coding operations that are considered in this paper. In this case, network coded packet is represented as a linear combination of the original source packets. In intra-session network coding, any network coded packet can be thus expressed in the form 
\begin{equation}
	y = {\bm a}{\bm x}^T = \sum \limits_{k = 1} ^ {N_s}a_kx_k,
	\label{eq:A}
\end{equation}
where $x_k$ is the $k$\textsuperscript{th} source packet, $a_k$ is the corresponding coding coefficient that has been randomly selected from the finite Galois field GF($q$) of size $q$ and $N_s$ denotes the number of source packets. The bold font is henceforth used to represent vectors. Thus, ${\bm a} = [a_1,\ldots,a_{N_s}]$ and ${\bm x}=[x_1,\ldots,x_{N_s}]$ are the vectors of coding coefficients and source packets, respectively.

In inter-session network coding, the operations performed in the network nodes consist in linear combinations of packets from different sources. In particular, if we denote as ${\bm x_s = [x_{s,1},\ldots,x_{s,N_s}]}$ and ${\bm a_s} = [a_{s,1},\ldots,a_{s,N_s}]$ the vectors of source packets and the corresponding coding coefficients of source $s$, an inter-session network coded packet can be written in the form 
\begin{equation}
	y = \sum \limits_{s = 1} ^ {S} {\bm a_s}{\bm x_s}^T = \sum \limits_{s = 1} ^ {S} 
	\sum \limits_{k = 1} ^ {N_s} a_{s,k}x_{s,k}
\label{eq:B}
\end{equation}
where $x_{s,k}$ represents the $k$\textsuperscript{th} source packet of source $s$ and $a_{s,k}$ is the corresponding coding coefficient, which is drawn from a finite Galois field GF($q$) according to a uniform distribution. $\mathcal{S}$ is the set of sources in the network, $S = |\mathcal{S}|$ is the total number of sources and $N_s$ is the number of original source packets of the $s$\textsuperscript{th} source. Depending on the sessions that are encoded together in each network coded packet, some of the vectors ${\bm a_s}$ can be all zero vectors. It is worth noting that if $\bm{a_s} = \bm{0}, \; \forall \; s\in \{1,2,...,S\}\backslash j$, ({\em i.e.}, only packets from source $j$ are combined to generate the packet $y$), Eq.~\eqref{eq:B} becomes equivalent to Eq.~\eqref{eq:A} and reduces to a case of intra-session network coding. In order to make the decoding process feasible, the network coded packets are augmented with a header of length $\sum _{s\in \mathcal{S}}N_s\log(q)$ bits, which contains the network coding coefficients. It becomes clear that the selection of the size of  GF($q$) results from a trade-off between the size of the overhead and the probability of generating linearly dependent packets, which evolves as $\mathcal{O}(1/q)$ \cite{RandomizedNC03}. 

At the clients, the decoding of a particular source is accomplished by means of Gaussian elimination on a set of packets when a sufficient number of intra- or inter-session network coded packets is collected. Specifically, upon receiving a network coded packet, the client stores the body message of the packet in the vector ${\bm y}$ and the packet header in the matrix ${\bm A}$. Thus, each row of the matrix ${\bm A}$ contains the coding coefficients of the corresponding encoded packet stored in ${\bm y}$ and encompasses all the transformations sustained by the packets as they travel through the network. The decoding of source $s$ is possible from any subset of rows of matrix ${\bm A}$ that (i) contains at least $N_s$ linearly independent rows with non-zero coefficient vectors at the position corresponding to source $s$ and (ii) is full rank, {\em i.e.}, the rank of the matrix formed by this subset of rows is equal to the number of variables with non-zero coding coefficients. Thus, if we denote as ${\bm A}^\prime$ the matrix formed by the rows that satisfy the above two conditions and as ${\bm y}^\prime$ the corresponding vector of encoded packets, then the packets of the $s$\textsuperscript{th} source can be recovered by solving the following system of equations
\begin{equation}
	{\bm y}^\prime = {\bm  A}^\prime {\bm x}^T,
\label{eq:C}
\end{equation}
where ${\bm x} = [ x_{1,1}, \ldots,x_{1,N_1},\ldots, x_{s,1}, \ldots,x_{s,N_s},\ldots]$ is the vector of the original source packets from all $S$ sources. It is worth mentioning that ${\bm x}$ cannot be decoded fully in Eq.~\eqref{eq:C}, and only the symbols of the $s$\textsuperscript{th} source are primarily recovered. 

To summarize, the decoding of source $s$ can be performed from intra-session network coded packets, inter-session network coded packets  or from a combination of packets from both categories. The last two cases lead to decoding additional packets along with the packets of source $s$. We assume that these unnecessary packets are simply dropped by the client, which is only interested by data from one source. By construction of the network coded packets, the coding coefficients' vectors contain many zeros. In an attempt to reduce the decoding complexity, the rows and the columns of the matrix ${\bm A}^\prime$ can be reorganized such that the resulting matrix is in the row echelon form. The vectors $ {\bm y}^\prime$ and ${\bm x}$ also have then to be reordered correspondingly, so that Eq.~\eqref{eq:C} yields a valid expression.

Even if inter-session network coding appears as a natural extension of intra-session network coding, the design of effective coding solutions is not trivial. While intra-session network coding has been shown sufficient to achieve the maximum rate in the case of a single multicast \cite{LinearNC03}, \cite{AlgebraicNC}, the optimal inter-session network coding solution is still an open probelm. The coexistence of several sources in the network poses challenges on the construction of the network codes. Random mixing of all the input packets in the intermediate network nodes obviously results in suboptimal performance since it would then be necessary to send information about all sources, blindly to all clients. However inter-session network coding can still be more efficient than solutions based on pure intra-session network coding \cite{TraskovISIT06}. Clearly, there exists a trade-off between the increase in the information content per packet ({\em i.e.}, the number of sessions that are mixed together), and the decoding delay experienced by the clients. In addition, some of the clients might even not have sufficient bandwidth to collect the necessary number of inter-session coded packets for decoding the source that they have subscribed to. In the following sections, we focus on the problem of identifying the cases where inter-session network coding is beneficial and we design coding strategies that are able to exploit efficiently the available network resources.


\section{Coding system description}
\label{sec:system}

We now describe in details the system that we consider in this paper. The network is modeled as a directed acyclic graph $\mathcal{G} =\left( \mathcal{V},\mathcal{E} \right)$, where $\mathcal{V}$ is the set of network nodes and $\mathcal{E}$ the set of network links. The links are characterized by a capacity $c_{ij}$, which is expressed in packets per second, and an average packet loss rate $\pi_{ij}$, where $(i,j)\in \mathcal{E}$ denotes the link between the nodes $i$ and $j$, with $i,j\in \mathcal{V}$. We also assume that the propagation delay on each link is negligible.

We assume that our network consists of a set of $S$ sources $\mathcal{S} = \{s\}$, $s = 1,2,...,S$, a set of $M$ network clients $\mathcal{C} = \{c\}$, $c=1,2,...,M$ and a set of intermediate network nodes $\mathcal I$. We thus have $\mathcal{V} = \mathcal{S}~\cup ~\mathcal{C} ~\cup ~\mathcal{I}$. An illustration of the system that we consider is depicted in Fig.~\ref{fig:network}. The sources simultaneously transmit uncorrelated information to the network clients. Each client is interested in receiving only one of the sessions. Prior to transmission by the sources, random linear combinations of the original packets that belong to the same sources are performed. This permits to generate packets of equal importance at each source and thus to alleviate the need for precise packet scheduling mechanisms. Moreover, it enhances the network resilience to packet losses, since any subset of $N_s$ linearly independent coded packets is sufficient for decoding. Finally, it provides high packet diversity to the system, which enables an easier and better exploitation of the network resources. 

Intermediate network nodes perform the task of storing, combining and forwarding packets to the next hop nodes in a push-based policy. In particular, intermediate nodes store the incoming packets, randomly combine them and forward the resulting packets on the output links, whenever there is a transmission opportunity. However upon receiving a packet, the nodes first examine whether this packet is innovative with respect to the packets stored in the nodes' buffers. A packet is considered innovative when it increases the rank of the system formed by the set of packets that exist in the node's buffer. In other words, a packet is classified as innovative when it cannot be generated by simply recombining the packets that are already stored in the node. Packets that are non-innovative are dropped immediately as they do not provide any novel information. The innovative packets are processed and stored in the nodes' buffers which are considered to be large enough to accommodate all the received packets. 

The percentage of the intra- and inter-session network coded packets that are forwarded by the nodes is determined by the coding policy. The design of effective coding policies is exactly the topic of this paper. In that perspective, we further need to define packet types and notation. Let us define as $\mathcal{T} = \{t\}$, $t=1,2,...,2^{S}-1$,  the set of possible packet types that can be generated in the network. Every packet type $t$ represents a particular combination of sources including intra-session network coded packets. With every packet type $t\in\mathcal{T}$ we associate a subset of packet types $\mathcal{T}_t \subseteq \cal{T}$ and a subset of sources $\mathcal{S}_t \subseteq \mathcal{S}$.  The elements of $\mathcal{T}_t$ are all the packet types that can be combined in order to form packets of type $t$. The subset $\mathcal{S}_t$ contains all the sources whose packets are combined in a network coded packet of type $t$. In addition, we define the subset $\mathcal{T}_{t,s} \subseteq \mathcal{T}_t$, which includes all the packet types in $\mathcal{T}_t$ that contain data from source $s$. 
\begin{table}
	\begin{center}
	\caption{Notations.\label{tab:notation}}
		\begin{tabular}{| c | p{6.9cm} |}
			\hline
			$c_{ij}$ & capacity of the link $(i,j)$ in packets/sec \\
			\hline
			$\pi_{ij}$ & packet loss rate on the link $(i,j)$ \\
			 \hline
			 $N_s$ &  number of packets in source $s$ \\
			 \hline
			 $\mathcal{T} $ & set of all possible packet types that can be generated in the network\\
			 \hline
			 $\mathcal{T}_t$ & set of all packet types that can be used to form packets of type $t$ \\
			 \hline
			  $\mathcal{S}_t$ & set of all sources that are combined to generate packets of type $t$ \\
			  \hline
			  $\mathcal{T}_{t,s}$ & set of all packet types that can be used to form packets of type $t$ and contain source $s$ \\
			  \hline
			  $\mathcal{A}_i$ & set of parent nodes of node $i$ \\
			  \hline
			  $\mathcal{D}_i$ & set of children nodes of node $i$ \\
			  \hline
			  $f_{ij}^t$ & flow rate of packets of type $t$ on link $(i,j)$ \\
			  \hline
			  $r_{ij}^t$ & innovative input flow rate of packets of type $t$ on link $(i,j)$ \\
			  \hline
			  $w_{ij}^t$ & probability of forwarding a packet of type $t$ on link $(i,j)$ \\
			  \hline
			  $P_c^s(k)$ & probability of decoding source $s$ with exactly $k$ packets at client $c$ \\
			  \hline
			  $p_c^t$ & probability of receiving a useful packet of type $t$ at client $c$ \\
 			  \hline
			  $d_c$ & average delay for receiving one packet at client $c$ \\
 			  \hline
			  $D_c^s$ & expected delay observed at client $c$ for decoding source $s$ \\
			  \hline
			  $\overline{D}$ & average expected delay \\
			  \hline
		\end{tabular}
	\end{center}
\end{table}
According to our model, every network node forwards to its neighbor nodes all or a subset of all the possible packet types, depending on the type of the incoming packets and the coding strategies in the nodes. Hence, the capacity $c_{ij}$ of the link $(i,j)$ is partitioned into several packet flows, each corresponding to a certain packet type $t$, and every packet flow is allocated a rate of $f_{ij}^t$ packets per second. Whenever a transmission opportunity occurs on the link ${(i,j)}$, the node selects randomly a packet type $t$ by sampling a probability distribution $\bm{w}_{ij}=\{w_{ij}^t\}$, $t \in \mathcal{T}$, which determines the relative number of packets of each type to be transmitted. In particular, $w_{ij}^t = {f_{ij}^t}/{c_{ij}}$ is the probability of sending a packet of type $t$ on the link $(i,j)$. Then, the node randomly combines the available packets to generate a packet of type $t$ that is sent on the output link. The network clients decode the network coded packets as described in Section \ref{sec:interNC}, and extract the information that is relevant to the session that they have subscribed to. The delay experienced by the clients for collecting a sufficient number of packets for decoding is driven by the number of innovative packets that they receive, which is a function of the flow rates and the network topology. 

Table \ref{tab:notation} summarizes the notation used throughout this paper.


\section{Decoding delay minimization}
\label{sec:ratealloc}

As we have seen above, the coding strategy implemented in the network nodes is key to the effective performance of inter-session network coding systems. We propose in this section a solution for minimizing the average decoding delay among the client population. This can typically be achieved by limiting the number of packets that each client has to decode. We first derive a model for estimating the expected delay at the clients required to decode the packets of the source that they have subscribed to. Our objective is to find the optimal coding policy in the network nodes defining the number of packets of each type that have to be generated and forwarded on the output links  so as to minimize the average expected decoding delay observed at the client nodes. We formulate an optimization problem and derive the optimal coding strategy using stochastic approximation algorithms. 

\subsection{Computation of the expected delay}
\label{sec:delay}

The time delay experienced at a client before it collects a decodable set of packets depends on the coding decisions implemented in the intermediate network nodes and the innovative rate of each packet type. The expected delay observed at a client node can be computed by estimating the average number of packets that the client receives before it is able to decode. To this end, let us assume that the client $c$ is interested in receiving data from source $s$ and denote as $D_c^s$ the average delay observed at client $c$ for receiving a sufficient number of packets in order to decode this data. This delay can be written as 
\begin{equation}
	D_c^s= d_c\sum \limits_{k = N_s}^{\infty} kP_c^s(k)
	\label{eq:J}
\end{equation}
where $k$ is the number of packets that client $c$ receives before being able to decode the information that it is interested in and $P_c^s(k)$ stands for the probability of decoding this information after receiving exactly $k$ packets. The constant $d_c$ represents the time needed for receiving one packet and we approximate it as  $$d_c =\frac{1}{\sum _ {n\in \mathcal{A}_c}c_{nc}}, $$ where $\mathcal{A}_c$ denotes the set of parent nodes of client $c$ and $c_{nc}$ represents the capacity of the network links serving the client $c$. 

It should be pointed out here that, for the sake of consistency, $k$ in Eq.~\eqref{eq:J} represents not only the packets, either innovative or not that reach client $c$, but also the time slots when no packet arrives at the client. In other words, whenever a packet is lost on the channel or a transmission opportunity is skipped by a parent node, we consider it as equivalent to receiving a useless packet. The minimum number of packets needed for decoding the source $s$ is equal to the size $N_s$ of the data. Hence, the probability of decoding with less packets than $N_s$ ({\em i.e.}, $k < N_s$) is equal to zero.  The upper limit of the summation in Eq.~\eqref{eq:J} is the maximum number of packets that a client may receive before it is able to decode. Theoretically, this number goes to infinity. In practice, however, there always exists a large enough number of packets $K_{max}$ for which the probability of decoding with less then $K_{max}$ packets is arbitrarily close to 1.

The probability $P_c^s(k)$ of decoding the packets of source $s$ at client $c$ with exactly $k$ packets is the probability of forming a full rank system upon receiving the $k$\textsuperscript{th} packet, but not earlier than that. This probability can be written as follows
\begin{equation}
	\begin{split}
		P_c^s(k) &= \sum_{t^* \in \mathcal{T}} p_c^{t^*}\Big\{ \sum_{k_1}\sum_{k_2}\dots \\
		&\sum_{k_{|\mathcal{T}|}} \binom{k-1}{k_1,k_2,\ldots,k_{|\mathcal{T}|}, (k-1-
		\sum\limits_{\scriptscriptstyle t\in \mathcal{T}}k_t)} \\ 
		 &\prod_{t\in\mathcal{T}}{(p_c^t)}^{k_t}{(1-\sum_{t\in\mathcal{T}}p_c^t)}^{k-1-
		\sum\limits_{\scriptscriptstyle t\in\mathcal{T}}k_t}\Big\}
	\end{split}
\label{eq:K}
\end{equation}
where  $k_t$ denotes the number of useful packets of type $t$ out of the total number of $k-1$ packets received by the client $c$. Since $P_c^s(k)$ is the probability of decoding source $s$ with exactly $k$ packets and not less than $k$, the range of values of $k_1, k_2,\ldots,k_{|\mathcal{T}|}$ can be computed so as to satisfy this condition. In particular, the $\sum_{t\in\mathcal{T}}k_t $ packets should contain a subset of packets that, when increased by the $k$\textsuperscript{th} packet, yields a full rank decodable system for the $s$\textsuperscript{th} source. The term $p_c^t$ represents the probability that a useful packet of type $t$ arrives at client $c$. $t^*$ denotes the type of the last, ({\em i.e.}, $k$\textsuperscript{th}) packet and $p_c^{t^*}$ is simply the probability that this packet is of type $t^*$. A packet of type $t$ is characterized as useful when it is innovative with respect to the packets that client $c$ has already received. Hence, this probability is written as
\begin{equation}
	p_c^t = \frac{\sum_{n\in\mathcal{A}_c}r_{nc}^t}{\sum_{n\in\mathcal{A}_c}c_{nc}},
\label{eq:L1}
\end{equation}
which defines the probability $p_c^t$ as the fraction of the total input innovative rate of packets of type $t$, $r_{ij}^t$, over the total input bandwidth. After combining Eqs~\eqref{eq:J} and \eqref{eq:K}, we obtain the following closed form expression for the expected decoding delay at client $c$ that decodes source $s$, as 
\begin{equation}
	\begin{split}
		D_c^s & = d_c\sum_{t^* \in \mathcal{T}} p_c^{t^*} \sum_{k_1}\sum_{k_2}\dots \\
		&\sum_{k_{|\mathcal{T}|}} \frac{(\sum_{ \scriptscriptstyle t\in \mathcal{T}}k_t + 1)!}
		{k_1!k_2!\ldots k_{|\mathcal{T}|}!}\frac{\prod\limits_{\scriptscriptstyle t\in\mathcal{T}}{(p_c^t)}^{k_t}}
		{\Big(\sum\limits_{\scriptscriptstyle t\in \mathcal{T}}p_c^t\Big)^{\sum\limits_{\scriptscriptstyle t\in \mathcal{T}}k_t + 2}}
	\end{split}	
	\label{eq:L2}
\end{equation}
The detailed development of Eq.~\eqref{eq:L2} is provided in the Appendix.

\subsection{Optimization of coding decisions}
\label{sec:probform}

We are now able to formulate the delay minimization problem that seeks for the optimal coding decisions at the network nodes and the corresponding allocation of rate among the different packet types. Specifically, we want to determine the probability distribution ${\bm w}_{ij} = \{w_{ij}^t\}$ according to which every node pushes packets on its output links, or equivalently, the flow rates $f_{ij}^t \geq 0$ for all the packet types and for all the network links. The optimization problem consists in minimizing the average expected delay $\overline{D}$ and can be written as follows
\begin{equation}
	\begin{split}
					&\quad\quad\quad \quad\underset{\{r_{ij}^t\}}{\operatorname{arg\,min}}\overline{D} = 
					 \underset{\{r_{ij}^t\}}{\operatorname{arg\,min}}\frac{1}{|\mathcal{C}|}\sum_{c \in \mathcal{C}}D_c^s \\
	&\mbox{s.t.} \\ \quad	& r_{ij}^t \geq 0, \; \forall t\in\mathcal{T}\\
					& \sum_{t\in\mathcal{T}}r_{ij}^t \leq c_{ij}(1-\pi_{ij}), \; \forall (i,j) \in \mathcal{E}\\
					&r_{ij}^t = 0, \; \mbox{if} \; \prod_{s \in \mathcal{S}_t}\Big(\sum_{t^\prime \in \mathcal{T}_{t,s}}
					\sum_{k \in \mathcal{A}_i}r_{ki}^{t^\prime}\Big) = 0,
					 \forall t\in\mathcal{T}, \;  \forall (i,j) \in \mathcal{E}\\
					&\sum_{t^\prime \in \mathcal{T}_{t,s}}r_{ij}^{t^\prime} \leq \sum_{t^\prime \in \mathcal{T}_{t,s}} 
					\sum_{k \in \mathcal{A}_i}r_{ki}^{t^\prime}, \; \forall t\in\mathcal{T},  \; 
					\forall s\in\mathcal{S}_t , \;\forall (i,j) \in \mathcal{E}\\
					&\sum_{t^\prime \in \mathcal{T}_{t,s}}\sum_{k \in \mathcal{A}_i}r_{ki}^{t^\prime} \leq 
					\sum_{l \in \mathcal{D}_s}r_{sl}, \; \forall t \in \mathcal{T}, \;\forall s\in\mathcal{S}_t, \; \forall i \in \mathcal{V}\backslash \mathcal{S}
	\end{split}
\label{eq:M}
\end{equation} 

The first two constraints of the optimization problem in Eq.~\eqref{eq:M} arise from the fact that the input innovative flow for every packet type is non-negative and the total input innovative flow on a link cannot exceed the effective bandwidth of the link, respectively. The third constraint states that the flows that cannot be generated due to the unavailability of some of the necessary component flow should not be allocated any rate. The fourth constraint states that,  for every packet type, all the components have an innovative flow on the output links that is upper bounded by the input innovative flow. Finally, the last constraint is similar to the fourth constraint except for the fact that the total input innovative rate is upper bounded by the total innovative rate provided by the sources. We illustrate the meaning of the two last constraints with the following example.

\begin{figure}[t]
	\begin{center}
		~\includegraphics[width=0.45\textwidth]{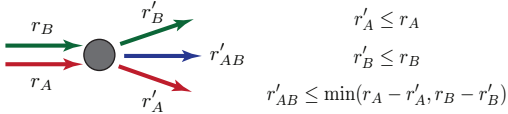}~
	\end{center}
	\caption{Illustration of the fourth constraint in the optimization problem of Eq~\eqref{eq:M}. The rate allocated to the combined flow does not exceed the minimum of the remaining packets in each input flow.}
	\label{fig:constraint3}
\end{figure}

\begin{example}
Let us consider a node in the network that receives innovative intra-session network coded packets from sources $A$ and $B$ at rates $r_A$ and $r_B$ respectively. The node is forwarding intra-session network coded packets of sources $A$ and $B$   and also combined packets of the two sources. This scenario is illustrated in Fig.~\ref{fig:constraint3}. If $r_A^\prime$, $r_B^\prime$ and $r_{AB}^\prime$ are respectively the innovative rates of the three types of flows delivered to the next hop nodes, then the third constraint in the optimization problem in Eq.~\eqref{eq:M} states that
\[
r_A^\prime \leq r_A
\]
\[
r_B^\prime \leq r_B 
\]
\[
r_{AB}^\prime \leq \min(r_A - r_A^\prime,  r_B -r_B^\prime)
\]
\end{example}

\noindent The above example illustrates that the maximum rate allocated to the flow of combined packets is upper-bounded by the minimum of the available innovative rates for each component flow. For example, if $r_A - r_A^\prime < r_B -r_B^\prime$, any additional rate allocated to the combined flow, such that $r_{AB}^\prime > r_A - r_A^\prime$ does not carry any novel information with respect to the source $A$. Therefore, it can rather be used for intra-session network coded packets from source $B$ with the same impact on the decodability. 

The minimization problem defined in Eq.~\eqref{eq:M} is in general non-convex and the number of variables increases with the number of sources and links in the network, which renders the search space huge. In order to find the optimal solution, we use the SPSA (Simultaneous Perturbation Stochastic Approximation) algorithm \cite{SPSASpall}, which is an efficient gradient based stochastic algorithm for finding a good approximation of the global optimum in multivariate non-convex optimization problems. With the solution of Eq.~\eqref{eq:M}, the flow rate for each packet type can be computed as $f_{ij}^t = \frac{r_{ij}^t}{\sum_{t\in\mathcal{T}}r_{ij}^t}c_{ij}$, and the allocation vector $\bm{w}_{ij}$ with $w_{ij}^t =\frac{f_{ij}^t}{c_{ij}}$ is given for all intermediate nodes. 


\section{Performance evaluation}
\label{sec:evaluation}

We now evaluate the performance of our inter-session network coding system with the proposed rate allocation technique. We compare the inter-session network coding scheme with a baseline intra-session network coding solution where network coding operations are performed across packets of the same session. In order to obtain the optimal rate allocation for the baseline scheme, we simply modify the optimization problem in Eq.~\eqref{eq:M} and restrict the coding operations to combinations of packets of the same session by setting the rate of the inter-session network coded flows to zero. 

\begin{figure*}[t]
	\begin{center}
			\subfloat[Topology 1]{\label{fig:topology1}\includegraphics[width=0.4\textwidth]{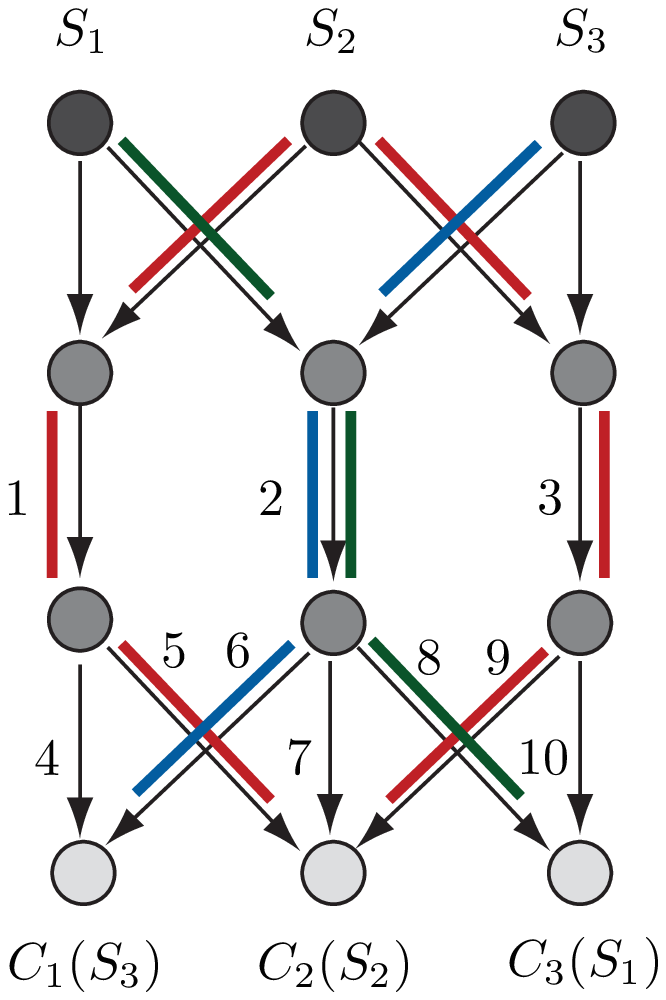}}~~~~~~~~~~~
			\subfloat[Topology 2]{\label{fig:topology2}\includegraphics[width=0.4\textwidth]{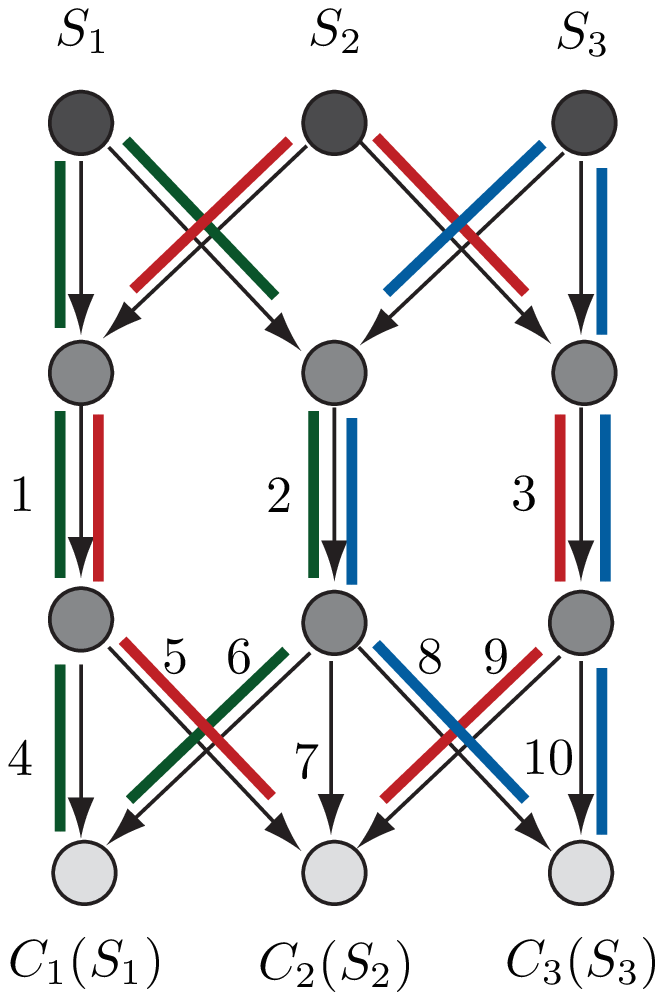}}
	\end{center}
	\caption{Examples of two network topologies with three concurrent unicast sessions. The paths connecting the clients with the source of their interest are highlighted with different colors. The notation $C_j(S_i)$ implies that the client $C_j$ requests data from the source $S_i$.}
	\label{fig:topology}
\end{figure*}

\begin{figure*}[t]
	\begin{center}
			\subfloat[Topology 1]{\label{fig:delay1}\includegraphics[width=0.4\textwidth]{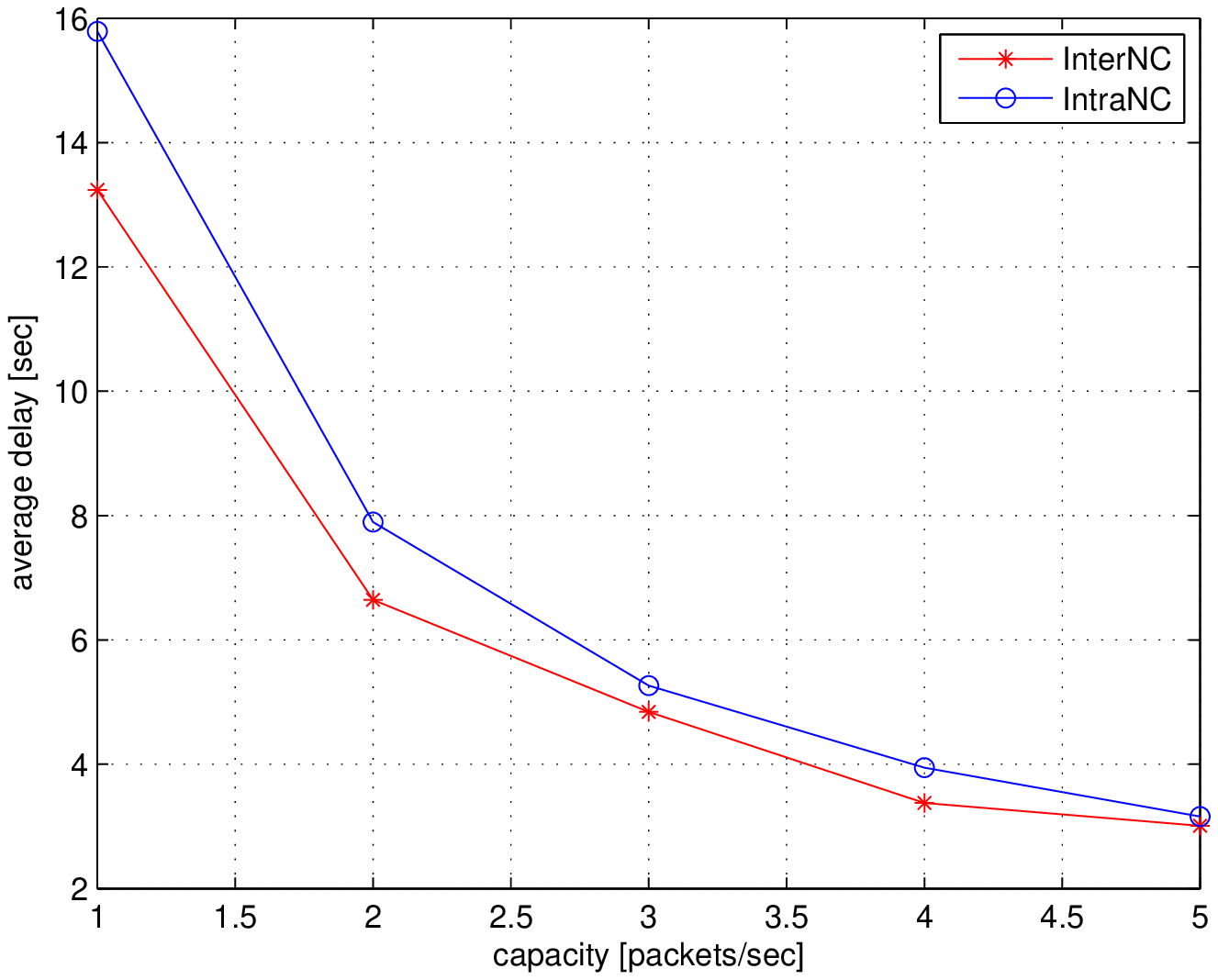}}~~~~~~~~~~~	
			\subfloat[Topology 2]{\label{fig:delay2}\includegraphics[width=0.4\textwidth]{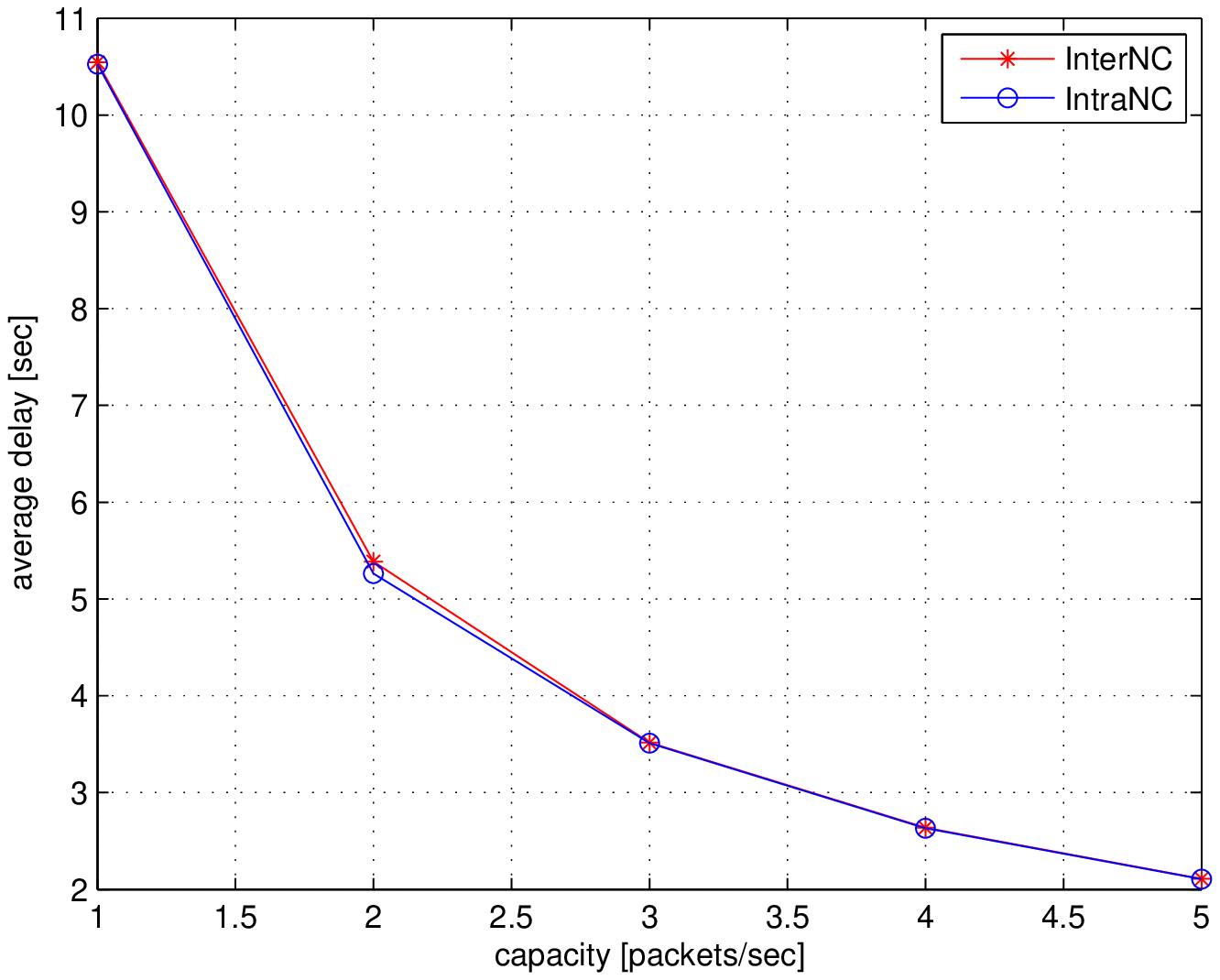}}
	\end{center}
	\caption{Comparison of the minimum average delay achieved with the optimal rate allocation for the proposed inter-session
	network coding scheme, and the baseline intra-session network coding scheme for the topologies illustrated in 
	 Fig.~\ref{fig:topology}.}
	\label{fig:delay}
\end{figure*}

\begin{figure*}[t]
	\begin{center}
			\subfloat[Topology 1]{\label{fig:ratealloc1}\includegraphics[width=0.45\textwidth]{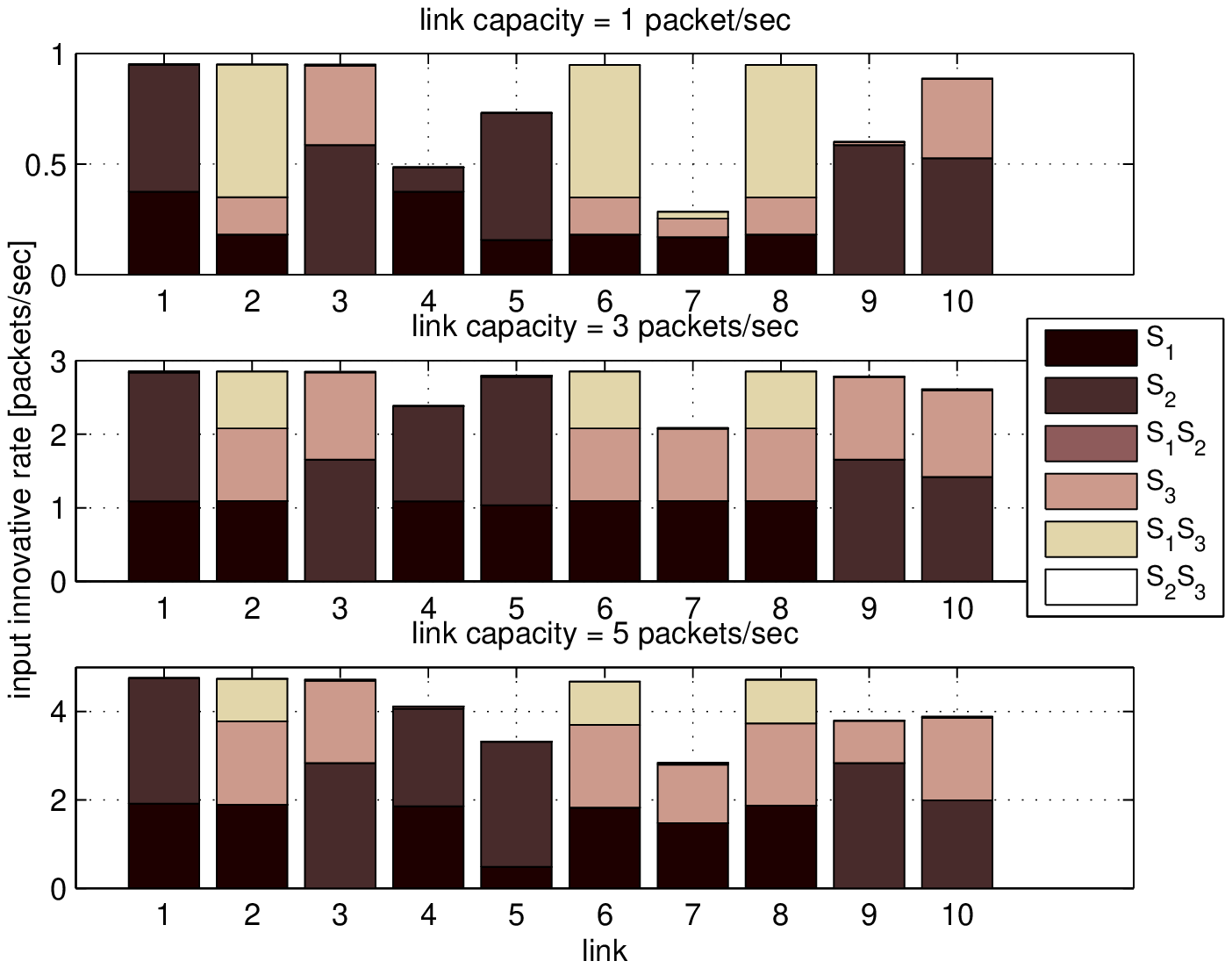}}~~~~~~~~~~~	
			\subfloat[Topology 2]{\label{fig:ratealloc2}\includegraphics[width=0.45\textwidth]{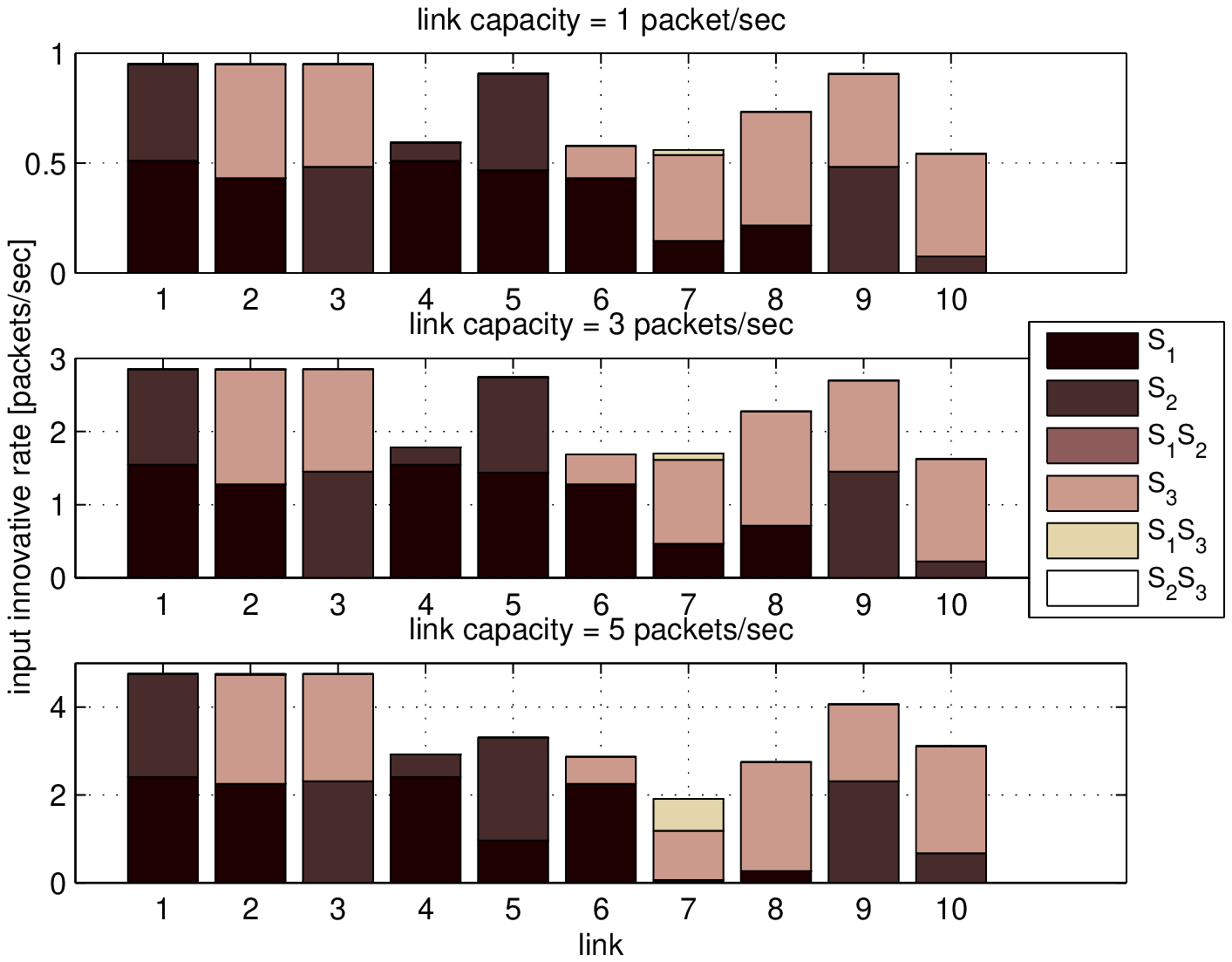}}
	\end{center}
	\caption{Optimal rate allocation of the innovative packet flows obtained with the proposed inter-session network coding scheme for the topologies illustrated in Fig.~\ref{fig:topology}.}
	\label{fig:rateallocation}
\end{figure*}

\begin{figure*}[t]
	\begin{center}
			\subfloat[Topology 1]{\label{fig:clientdelay1}\includegraphics[width=0.45\textwidth]{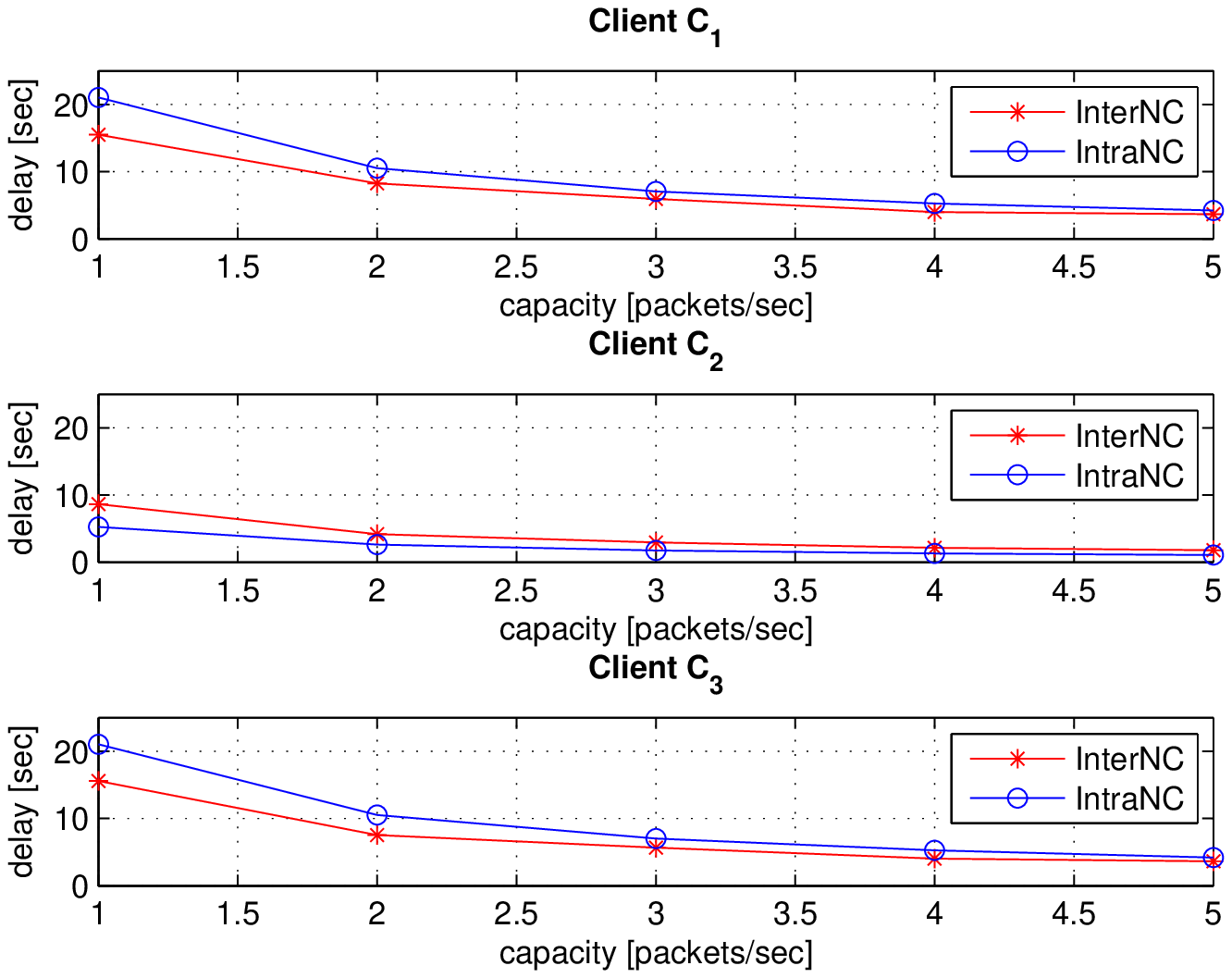}}~~~~~~~~~~~
			\subfloat[Topology 2]{\label{fig:clientdelay2}\includegraphics[width=0.45\textwidth]{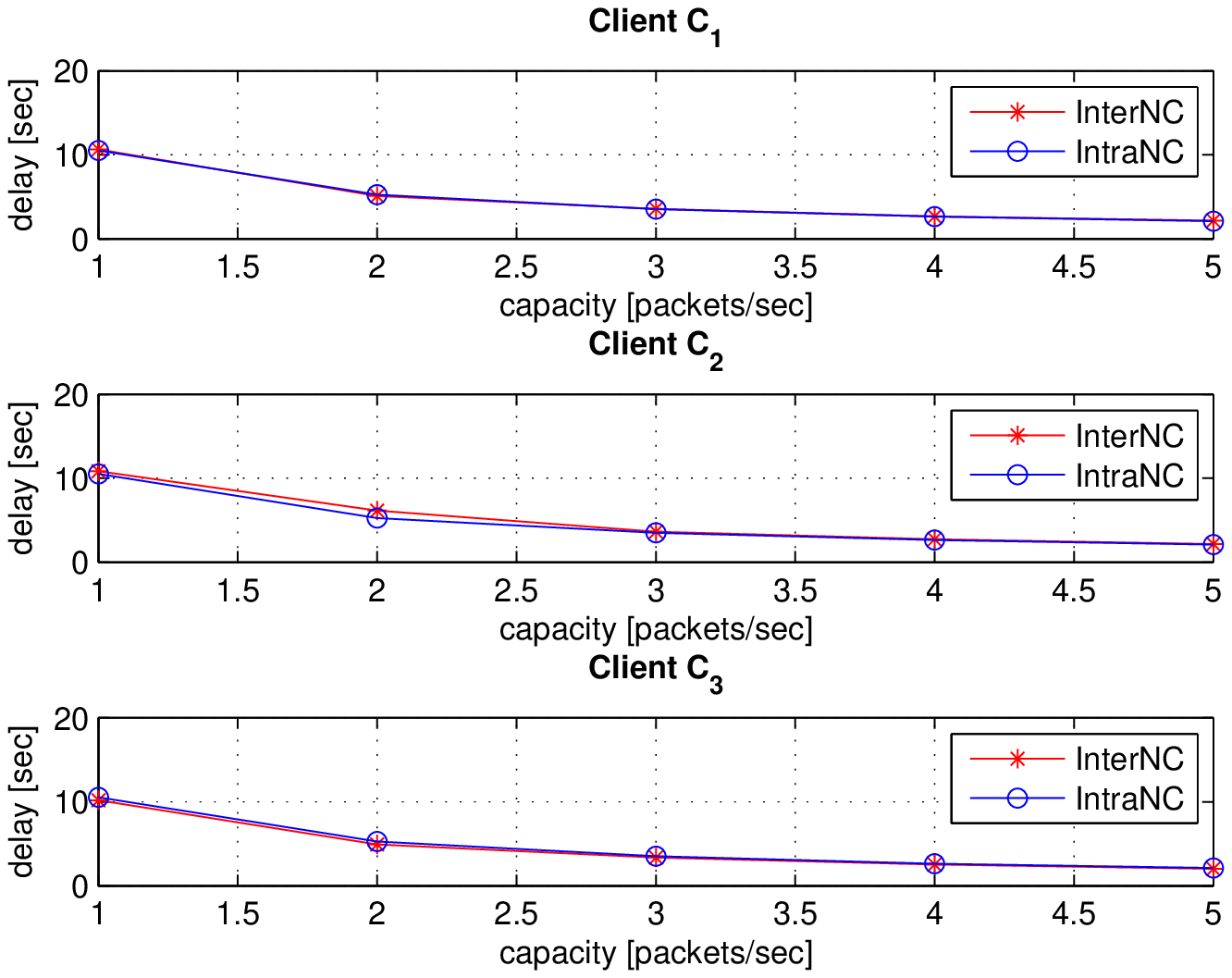}}
	\end{center}
	\caption{Comparison of the decoding delay experienced by the network clients for the proposed inter-session network coding scheme and the baseline intra-session network coding scheme for the topologies illustrated in Fig.~\ref{fig:topology}.}
	\label{fig:clientdelay}
\end{figure*}

To validate the correctness of our model, we provide results for two small size network topologies illustrated in Fig.~\ref{fig:topology}, though our findings can be extended to any arbitrary topology. Both topologies that we consider consist of $S = 3$ source nodes, $M = 3$ client nodes and 6 intermediate nodes. Every client is interested in receiving packets of only one of the available sources, as shown in Fig.~\ref{fig:topology}. The original packets are encoded at the sources with randomized linear coding to increase the symbol diversity and then forwarded to the next hop nodes. The size of the GF for coding operations is set to $q= 256$. We assume that the size of the data is $N_s = 10$ packets for all sources and every source node is transmitting  with a fixed rate of 3 packets/sec over each outgoing link. The intermediate nodes randomly forward network coded packets according to the rate allocation obtained from the solution of the optimization problem in Eq.~\eqref{eq:M}. Finally, the clients decode the source of their interest upon collecting a decodable set of packets.

In Fig.~\ref{fig:delay1} we present the average expected delay experienced by the network clients as a function of the links' bandwidth for the topology depicted in Fig.~\ref{fig:topology1}. The capacity of the links 1-10 varies in the interval $[1,5]$ packets/sec and the packet loss rate is set to $5\%$. The proposed inter-session network coding scheme outperforms the baseline network coding scheme in the whole range of capacity values. Higher gain in terms of delay is observed in the presence of heavy bottlenecks, {\em i.e.}, when the ratio of the input to the output bandwidth in the first hop helper nodes is high. The gain drops gradually as this ratio decreases. The performance of the inter-session scheme approaches the one of the intra-session network coding scheme, as the bandwidth becomes sufficient to transmit complete intra-session network coded sources.

For the same network settings, Fig.~\ref{fig:ratealloc1} illustrates the optimal rate allocation of innovative packet flows obtained for the proposed inter-session network coding scheme for three link capacity values, namely 1, 3 and 5 packets/sec. We can observe that the optimal network coding strategy is to forward combined packets of sources $S_1$ and $S_3$ on link 2 and to utilize some bandwidth on links 1 and 3 to transmit intra-session network coded packets that are used to decode the inter-session network coded packets. Thus, the system transmits on links 4 and 10, some  packets that are not explicitly useful for clients $C_1$ and $C_3$, but that are exploited to decode the inter-session network coded packets. This leads to reduced decoding delay for clients $C_1$ and $C_3$ at the expense of slightly increasing the decoding delay for client $C_2$. Indeed, part of the bandwidth that is utilized to provide packets of source $S_2$ to client $C_2$ in the intra-session network coding scheme becomes dedicated to flows of packets that are useful to decode inter-session network coded packets. As the ratio of the input bandwidth over the output bandwidth decreases, the proportion of the inter-session network coded  packets also drops and the optimal coding strategy converges to an intra-session network coding solution.

The decoding delay for each client is finally depicted in Fig.~\ref{fig:clientdelay1} for the same network settings. We can see that for the whole range of capacity values, the clients $C_1$  and $C_3$ have lower delay with inter-session network coding, whereas client $C_2$ experiences a slightly higher delay. This yields an overall gain in the average decoding delay over the network. This is due to the fact that some of the network clients have very scarce resources and are significantly affected by the network bottlenecks when only intra-session network coding is implemented. For example, (see Fig.~\ref{fig:topology1}), the client $C_2$ has two paths that are fully dedicated to its source of interest. However, the clients $C_1$ and $C_3$ each have only one path that share a common segment. Hence, the performance of the intra-session coding solution is limited by the bottleneck that is created by the overlapping paths. Moreover, the intra-session network coding strategy fails to fully utilise the network resources as some of the links cannot forward useful packets and thus some capacity remains unexploited. However, the overlap of the two paths creates opportunities for inter-session network coding. When packet combinations of different sources are allowed, the network links can be utilized to supply useful packets to the other clients in the network.  This permits to better exploit the available capacity and to redistribute the existing network resources more fairly among the clients. Hence, by slightly penalizing client $C_2$, the system manages to achieve a better average performance. As discussed earlier, the advantage of inter-session network coding decreases as the bandwidth increases to a value that is sufficient for transmitting each source at their source rate.

We perform the same experiments on a different topology illustrated in Fig.~\ref{fig:topology2}. As previously, the capacity of the links varies in the range $[1,5]$ packets/sec and the packet loss rate is set to $5\%$. From Fig.~\ref{fig:delay2} we observe that the performance of the proposed inter-session network coding scheme coincides with the performance of the baseline intra-session network coding scheme. This is due to the fact that the performance in this case is mostly driven by the available bandwidth rather than by the coding scheme. In particular, every client has two paths to the source and both paths overlap with a path of the other sources. Due to this symmetry in the network topology, the minimum delay can be achieved by allocating equal amounts of bandwidth to each of the sources and cannot be further reduced by inter-session network coding. In other words, the network does not create any opportunities for packet mixing across different sources. In fact, as depicted in Fig.~\ref{fig:ratealloc2}, the optimal rate allocation in this case contains only intra-session network coded flows. Interestingly, our scheme is however generic and includes the pure intra-session network coding scheme as a potential solution. The above results are further supported by the decoding delay experienced by each client, as presented in Fig.~\ref{fig:clientdelay2}. We can see that both schemes provide the same delay for every client, and that it cannot be improved by combining packets from different sources.

\begin{figure}[t]
	\begin{center}
		~\includegraphics[width=0.45\textwidth]{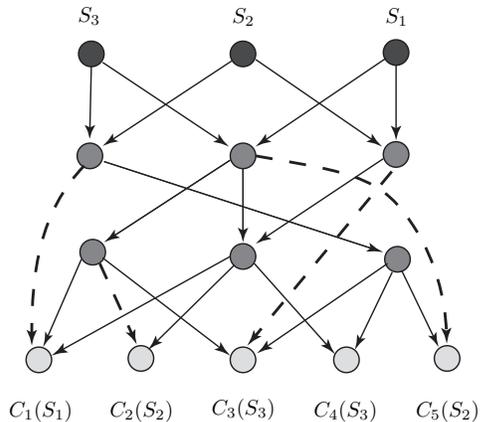}
	\end{center}
	\caption{Irregular multicast topology with three sources and five clients. The dashed line links have a fixed capacity of 10 packets/sec. The notation $C_j(S_i)$ implies that the client $C_j$ requests data from the source $S_i$. }
	\label{fig:remedytopology}
\end{figure}

In a third set of experiments, we consider the topology depicted in Fig.~\ref{fig:remedytopology} that consists of $S = 3$ sources, $M = 5$ clients and 6 helper nodes. The capacity of the sources' output links and the clients' input links is set to 30 packets/sec.  The links that are represented with dashed lines have a capacity fixed at 10 packets/sec. The capacity of the rest of the links varies in the interval $[5,30]$ packets/sec. We observe that the links in dashed lines cannot provide packets that are directly useful for clients if only routing or intra-session network coding are implemented, since they do not lie on a path connecting the clients with the sources that they request.  However, these links are helpful for delivering network coded packets that facilitate decoding at the clients when inter-session network coding is implemented in the intermediate nodes. 

\begin{figure}[t]
	\begin{center}
		\subfloat[]{\label{fig:delayremedy}\includegraphics[width=0.45\textwidth]{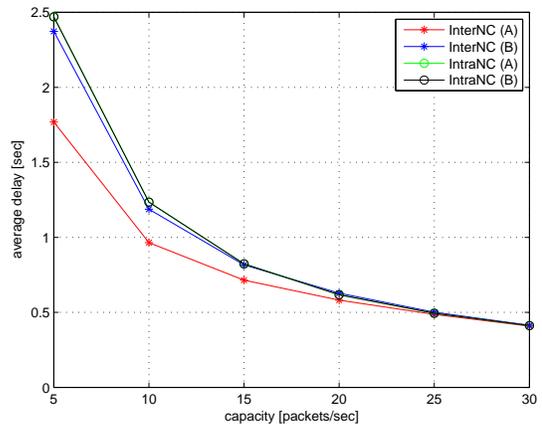}} \\
		\subfloat[]{\label{fig:rateallocremedy}\includegraphics[width=0.45\textwidth]{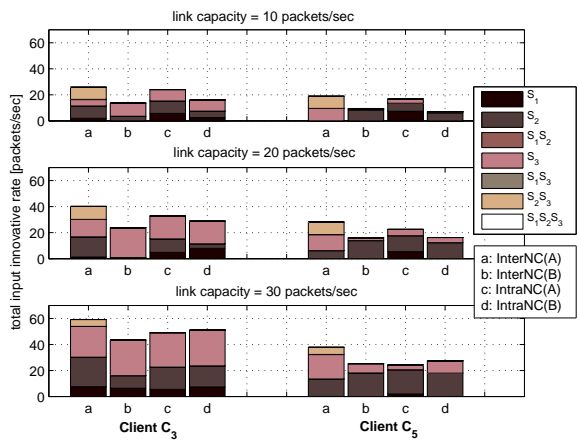}}
	\end{center}
	\caption{(a) Average expected delay experienced by the network clients and (b) optimal allocation of the input innovative rate for clients $C_3$ and $C_5$, for the topology depicted in Fig.~\ref{fig:remedytopology}, when intra- or inter-session network coding is deployed. The cases A and B correspond respectively to the addition, or not, of the dashed line links.  }
	\label{fig:remedy_delay_rate}
\end{figure}

Fig.~\ref{fig:delayremedy} illustrates the improvement in the average expected delay experienced by the network clients when the dashed line links are added to the network. It confirms that the addition of these links creates opportunities for inter-session network coding and eventually reduces the decoding delay. On the other hand, it is clear that the addition of these links cannot enhance the performance of the system when inter-session network coding is not enabled. When only intra-session network coding is implemented, the average expected delay in the presence of dashed line links is identical to the one achieved without these links (the green line in Fig.~\ref{fig:delayremedy} coincides with the black line). This confirms the fact that the dashed line links cannot provide packets that are explicitly useful for clients. Finally, we can observe that without the dashed line links, the proposed inter-session network coding scheme performs close to intra-session network coding solution in the topology of Fig.~\ref{fig:remedytopology}.

The above conclusions are supported by the rate allocation that achieves the minimal average decoding delay. Fig.~\ref{fig:rateallocremedy} shows the optimal allocation of the input innovative rate for clients $C_3$ and $C_5$ for increasing values of capacities. The schemes under comparison are the proposed inter-session network coding scheme and the intra-session network coding scheme. We can notice that for the pure intra-session network coding scheme, the introduction of the dashed line links does not change the optimal rate allocation solution. For the inter-session network coding scheme, when there are no dashed line links, the optimal solution is very close to the intra-session network coding solution. We can thus conclude that in this particular topology there do not exist many opportunities for inter-session network coding in the absence of dashed line links. However, we can see that the addition of these links leads to a different rate allocation solution where significant amounts of the available bandwidth are allocated to inter-session network coded flows. In particular, we can observe that clients $C_3$ and $C_5$ benefit from the combination of sources $s_2$ and $s_3$; they also receive intra-session network coded packets from sources $s_2$ and $s_3$ respectively, so that the decoding is facilitated.  Moreover, we can see that as the capacity of the bottleneck links increases, the rate of combined flows diminishes. This is in accordance with our intuition and the above findings. When the bandwidth is sufficient, all the packets can be provided as intra-session network coded packets.

\section{Conclusions }
\label{sec:conclusions}

We have proposed a delay minimal solution for rate allocation in multihop networks. Our scheme minimizes the average expected delay experienced by the network clients by allowing intermediate network nodes to transmit packets that are combinations of different sources. Specifically, we determine the optimal allocation of the intra- and inter-session network coded flows in the network nodes. We show that inter-session network coding achieves a better exploitation of the available network resources and offers significant gains in terms of decoding delay in networks with heavy bottlenecks. In less favorable topologies, inter-session network coding performs close to pure intra-session network coding as the latter is essentially a special case of inter-session network coding in our proposed system. The results of our simulations indicate that inter-session network coding can potentially introduce gains in terms of decoding delay to networks with heterogeneous clients, {\em i.e.}, clients with different access to the network resources. However, the full characterization of networks where inter-session network coding is superior to intra-session network coding still remains an open question. Moreover, though the proposed method is generic and can be applied to an arbitrary number of sources, the increased complexity of the solution prevents its utilization in real time applications. Our future work will focus on the design of distributed low-complexity inter-session network coding algorithms that would allow to take the optimal coding decisions in the network nodes in real time.


\appendix
\label{sec:appendix}
 
Let us consider a discrete random variable $k$ with a probability mass function (pmf) given by $P_c^s(k)$ as defined in Eq.~\eqref{eq:K}. The mean value of $k$ is then equal to 
\begin{equation}
	\begin{split}
		E[k]  &= \sum_{t^* \in \mathcal{T}}p_c^{t^*}  \sum_{k_1}\sum_{k_2}\dots \\
		\sum_{k_{|\mathcal{T}|}} &\frac{(\sum_{ \scriptscriptstyle t\in \mathcal{T}}k_t + 1)!}
		{k_1!k_2!\ldots k_{|\mathcal{T}|}!}\frac{\prod\limits_{\scriptscriptstyle t\in\mathcal{T}}{(p_c^t)}^{k_t}}
		{\Big(\sum\limits_{\scriptscriptstyle t\in \mathcal{T}}p_c^t\Big)^{\sum\limits_{\scriptscriptstyle t\in \mathcal{T}}k_t + 2}}
	\end{split}
	\label{eq:O}
\end{equation}

In more details, the mean value of $k$ is given by 
 \begin{equation}
 	\begin{split}
		&E[k] = \sum \limits_{k = N_s}^{\infty} kP_c^s(k)=
	  	\sum \limits_{k = N_s}^{\infty} k\sum_{t^* \in \mathcal{T}}p_c^{t^*}  \Big\{ \sum_{k_1}\sum_{k_2}\dots \\
		&\sum_{k_{|\mathcal{T}|}} \binom{k-1}{k_1,k_2,\ldots,k_{|\mathcal{T}|}, (k-1-\sum\limits_{\scriptscriptstyle t\in \mathcal{T}}k_t)} \\
		&\prod_{t\in\mathcal{T}}{(p_c^t)}^{k_t}
		{(1-\sum_{t\in\mathcal{T}}p_c^t)}^{k-1-\sum\limits_{\scriptscriptstyle t\in\mathcal{T}}k_t}\Big\}
	\end{split}
	\label{eq:P}
\end{equation}
Now, any set of values $\{k_1, k_2, \ldots, k_{|\mathcal{T}|}\}$ that leads to successful decoding with exactly $k$ packets, the last packet being of type $t^*$, also leads to successful decoding with $k+1,k+2,k+3$ or more packets and the last packet of type $t^*$. Thus, we can exchange the order of summations and rewrite Eq.~\eqref{eq:P} as 
\begin{equation}
	\begin{split}
		&E[k]  = \sum_{t^* \in \mathcal{T}} p_c^{t^*}\Big\{\sum_{k_1}\sum_{k_2}\dots 
		 \sum_{k_{|\mathcal{T}|}} \\
		& \sum \limits_{k =\sum\limits_{ \scriptscriptstyle t\in \mathcal{T}}k_t + 1}^{\infty}
		 k\frac{(k-1)!}{k_1!k_2!\ldots k_{|\mathcal{T}|}!(k-1-\sum\limits_{\scriptscriptstyle t\in \mathcal{T}}k_t)!}\\
		& \prod_{t\in\mathcal{T}}{(p_c^t)}^{k_t}
		{(1-\sum_{t\in\mathcal{T}}p_c^t)}^{k-1-\sum\limits_{\scriptscriptstyle t\in\mathcal{T}}k_t}\Big\}=\\
		 &\sum_{t^* \in \mathcal{T}}p_c^{t^*} \Big\{\sum_{k_1}\sum_{k_2}\dots \sum_{k_{|\mathcal{T}|}} 
		\frac{(\sum_{ \scriptscriptstyle t\in \mathcal{T}}k_t )!
		}{k_1!k_2!\ldots k_{|\mathcal{T}|}!}\frac{\prod\limits_{\scriptscriptstyle t\in\mathcal{T}}{(p_c^t)}^{k_t}}{\Big(\sum\limits_{\scriptscriptstyle t\in \mathcal{T}}p_c^t\Big)^{\sum\limits_{\scriptscriptstyle t\in \mathcal{T}}k_t + 1}} \\
		 &\Big(\sum \limits_{k =\sum\limits_{ \scriptscriptstyle t\in \mathcal{T}}k_t + 1}^{\infty}
		 k\frac{(k-1)!}{(\sum\limits_{ \scriptscriptstyle t\in \mathcal{T}}k_t )!(k-1-\sum\limits_{\scriptscriptstyle t\in \mathcal{T}}k_t)!} \\
		& \Big(\sum\limits_{\scriptscriptstyle t\in \mathcal{T}}p_c^t\Big)^{\sum\limits_{ t\in \mathcal{T}}k_t + 1}
	{\Big(1-\sum_{t\in\mathcal{T}}p_c^t\Big)}^{k-1-\sum\limits_{\scriptscriptstyle t\in\mathcal{T}}k_t}\Big)\Big\}
	\end{split}
\label{eq:Q}
\end{equation}
 The summation term with respect to $k$ in Eq.~\eqref{eq:Q}  represents the mean value of the negative binomial distribution NB($r,p$) with parameters $r = \sum\limits_{ \scriptscriptstyle t\in \mathcal{T}}k_t + 1$ and $p = \sum\limits_{\scriptscriptstyle t\in \mathcal{T}}p_c^t$ and is equal to 
 \begin{equation}
\mu = \frac{r}{p} = \frac{\sum\limits_{ \scriptscriptstyle t\in \mathcal{T}}k_t + 1}{\sum\limits_{\scriptscriptstyle t\in \mathcal{T}}p_c^t}
\label{eq:R}
 \end{equation}
 Combining Eqs~\eqref{eq:Q} and \eqref{eq:R}, we obtain the result in Eq.~\eqref{eq:O}. Finally, Eq.~\eqref{eq:L2} can be obtained by substituting the term $ \sum \limits_{k = N_s}^{\infty} kP_c^s(k)$ in Eq.~\eqref{eq:J}  with Eq.~\eqref{eq:O}.

\balance
\bibliographystyle{IEEEtran}
\bibliography{netcoding}

\end{document}